\documentclass[letterpaper, 10 pt]{ieeeconf}
\IEEEoverridecommandlockouts

\usepackage{cite}
\usepackage{amsmath,amssymb,amsfonts}
\usepackage{soul}
\usepackage{algorithmic}
\usepackage{graphicx}
\usepackage{textcomp}
\usepackage{xcolor}
\def\BibTeX{{\rm B\kern-.05em{\sc i\kern-.025em b}\kern-.08em
    T\kern-.1667em\lower.7ex\hbox{E}\kern-.125emX}}

\usepackage{subcaption} 
\begin{document}

\title{Optimal Path Planning for Wheel Loader Automation Enabled by Efficient Soil–Tool Interaction Modeling
\thanks{The authors are with the Department of Mechanical and Aerospace Engineering, University of California Davis, Davis, CA 95616 USA (e-mail: abdolmohammadi@ucdavis.edu; nmojahed@ucdavis.edu;
bravani@ucdavis.edu ; snazari@ucdavis.edu).
This work was supported in part by Komatsu Ltd. The authors gratefully
acknowledge their financial support.}
}

\author{Armin Abdolmohammadi, Navid Mojahed, Bahram Ravani and Shima Nazari}

\maketitle
\thispagestyle{empty}
\pagestyle{empty}

\begin{abstract}
Earthmoving operations with wheel loaders require substantial power and incur high operational costs. This work presents an efficient automation framework based on the Fundamental Earthmoving Equation (FEE) for soil–tool interaction modeling. A reduced-order multi-step parameter estimation method guided by Sobol’s global sensitivity analysis is deployed for accurate, online excavation force prediction. An optimal control problem is then formulated to compute energy-efficient bucket trajectories using soil parameters identified in the previous digging cycle. High-fidelity simulations in Algoryx Dynamics confirm accurate force prediction and demonstrate 15-40\% energy savings compared to standard paths. The total computation time is comparable to a single digging cycle, highlighting the framework’s potential for real-time, energy-optimized wheel loader automation.

\end{abstract}


\section{Introduction}

Autonomous excavation is increasingly studied in construction and mining. The bucket–filling phase of wheel–loader operation remains the most challenging, as it integrates complex soil–tool interactions with uncertain material properties and high energy use \cite{gottschalk2018test,haas2025combined}. These challenges motivate a pipeline for efficient bucket motion planning in soil enabled by soil parameter estimation. 

Bucket loading path planning has evolved along several key directions. Early work used “operator templates” from expert demonstrations and compared their efficiency in simulation \cite{filla2014_compare_trajectory}. In parallel, physics-based optimization then used dynamic programming on simplified loader models to minimize fuel/energy over full cycles, but at a substantial offline computation cost \cite{FRANK2018}. Simulation-driven design was used to explore large grids of trajectories and chart productivity–energy trade-offs across soils \cite{Aoshima_2021,chen2022}. More recently, data-driven methods have reproduced expert behavior via imitation learning and learned policies with reinforcement learning \cite{halbach2019_Pile_Loading_Controller,Eriksson2024,azulay2021_scooping_control}.

Despite their merits, these approaches often rely on extensive offline training or exhaustive scenario sweeps. Their performance is highly dependent on site-specific data, and transferring across soil types and stockpile shapes typically demands new experiments or large-scale simulation batches. A complementary direction is to use a physics-based planner grounded in analytical soil–tool interaction models. The Fundamental Earthmoving Equation (FEE) developed by Reece~\cite{reece1964paper} and refined by McKyes~\cite{mckyes1985_book} is widely adopted, providing estimates of excavation resistance based on soil properties and tool geometry. FEE is particularly attractive because it is analytical, lightweight, and interpretable.

FEE has been used for path planning \cite{yao2023bucket}, real-time control \cite{egli2022soil}, and hybrid physics–ML schemes \cite{yu2023line}, though it relies on soil parameters unknown on site. Prior work estimated these parameters via Newton–Raphson \cite{tan2005online}, while in our previous study \cite{abdolmohammadi2025data} we introduced a multi-stage constrained optimization method to identify them from excavation data, verified with Algoryx simulations \cite{algoryx2024} within a digital twin framework \cite{karanfil2025developing}. Data-driven approaches such as \cite{abtahi2025multi} could then be leveraged for path following, complementing the physics-based framework presented here.

This letter extends our framework in two directions: (i) we perform a global sensitivity analysis to construct Reduced–Order Models (ROMs), decreasing the dimensionality of the nonlinear, multi-stage soil parameter identification and accelerating convergence; and (ii) we embed the calibrated FEE model in an optimal path planner that minimizes cumulative excavation power subject to actuation, geometric, and payload constraints. These directions close the loop from excavation force measurement to soil parameter identification to energy–optimal bucket loading trajectories. 

\section{Methodology}
\label{sec:methodology}

\begin{figure*}[htbp]
    \centering
    \begin{subfigure}{0.23\textwidth}
        \centering
        \includegraphics[width=\textwidth]{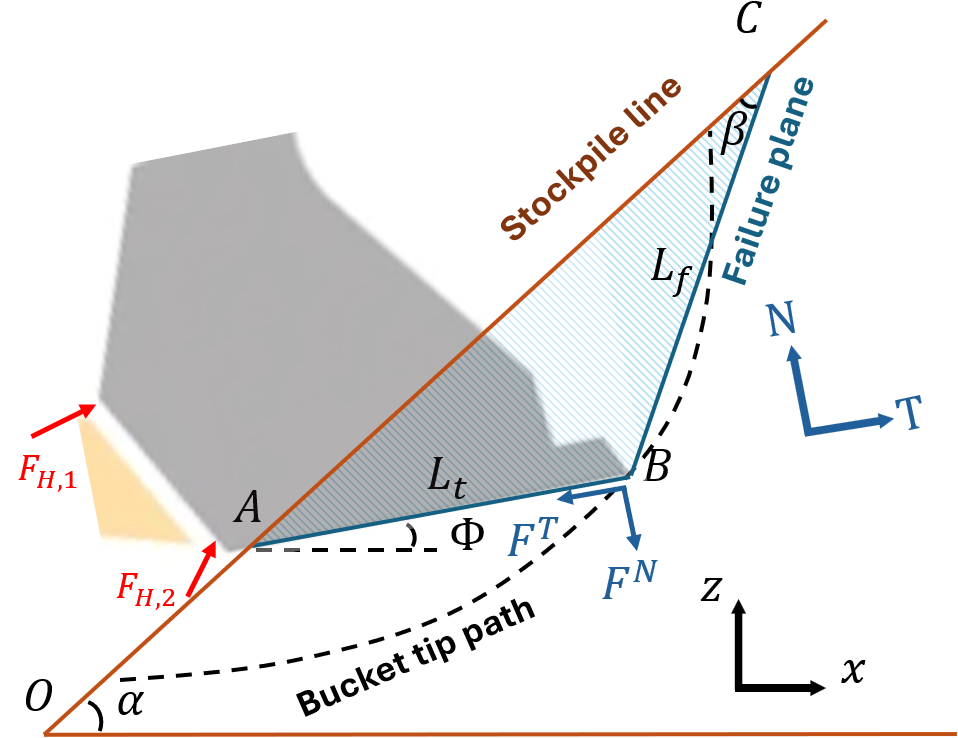}
        \caption{}
        \label{fig:bucketpath}
    \end{subfigure}
    \begin{subfigure}{0.20\textwidth}
        \centering
        \includegraphics[width=\textwidth]{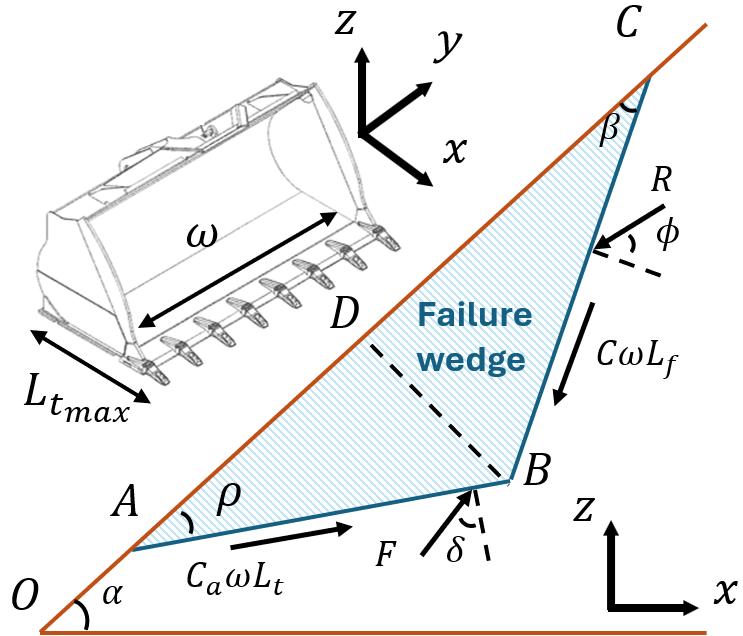}
        \caption{}
        \label{fig:FEE}
    \end{subfigure}
    \begin{subfigure}{0.24\textwidth}
        \centering
        \includegraphics[width=\linewidth]{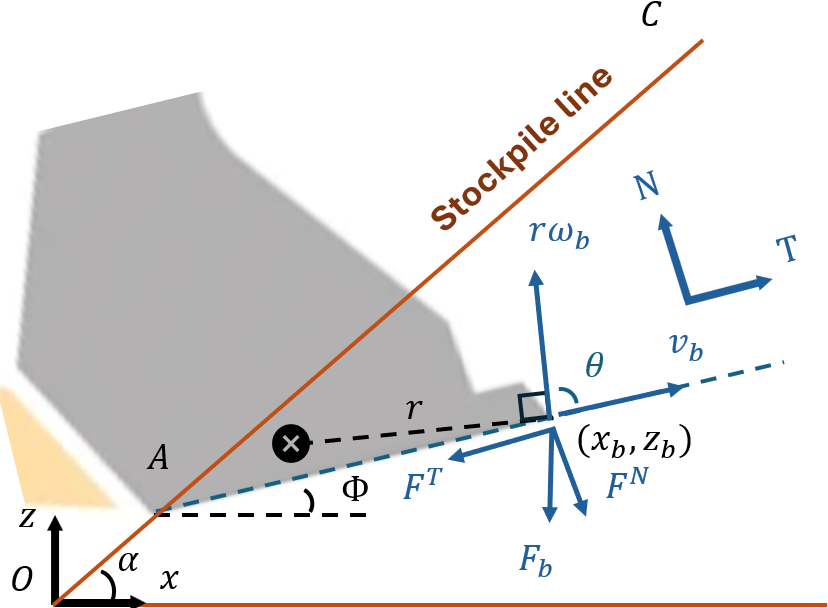}
        \caption{}
        \label{fig:MPPP}
    \end{subfigure}
    \begin{subfigure}{0.21\textwidth}
        \centering
        \includegraphics[width=\linewidth]{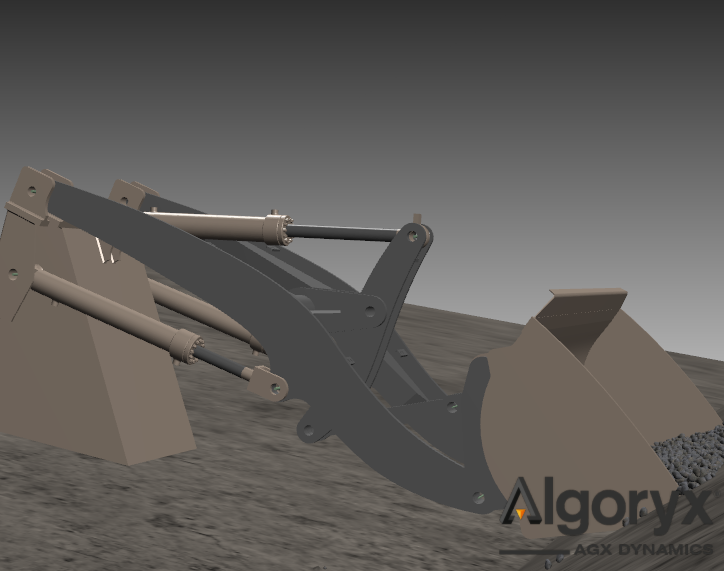}
        \caption{}
        \label{fig:Algoryx}
    \end{subfigure}
    \caption{(a) Bucket trajectory and failure plane $BC$. (b) Free-body diagram of the failure wedge with forces in the FEE model. (c) Bucket motion and soil states used in the path planner. (d) High fidelity model of the wheel loader in Algoryx.}
    \label{fig:results_multi_stage}
\end{figure*}

The soil–bucket interaction is modeled using the FEE framework. This model has been applied to load analysis in excavators~\cite{corke2000}, wheel loader design~\cite{worley2008}, and trajectory optimization in predictive control frameworks~\cite{yao2023bucket}. This framework discretizes the bucket motion in time and evaluates forces along the engaged blade segment at each step. As shown in Fig.~\ref{fig:bucketpath}, the bucket tip follows a dashed path in the global $x-z$ plane. Its orientation is given by the tilt angle $\Phi$, while the stockpile surface is idealized as the line $\overline{OC}$ with slope $\alpha$. The portion of the blade in contact with soil, $\overline{AB}$ with a length of $L_t$, experiences resistive forces resolved into tangential $F^T$ and normal $F^N$ components in the local $T-N$ frame. 

Let $F$ be the net frictional and normal force introduced by the FEE model that is applied to the soil wedge. Fig~\ref{fig:FEE} illustrates the free-body diagram of the failure wedge, where line $\overline{BC}$ denotes the failure plane. Force $F$ is applied at the external friction angle $\delta$, the soil reaction $R$ at the internal friction angle $\phi$, the cohesive force $Cw L_f$ along the failure plane of length $L_f$ with $C$ being the cohesion coefficient, and the adhesive force $C_a w L_t$ along the blade surface with $C_a$ being the adhesion coefficient. The lengths $L_t$ and $L_f$ evolve as the blade advances. The forces acting on the bucket blade can be written as:
\begin{align}
F^{T} &= w bP + F\sin\delta + C_a w L_t, \label{eq:FT}\\
F^{N} &= F\cos\delta, \label{eq:FN}
\end{align}
 where  $w$ is the bucket width. The normal penetration pressure \( P \), which builds up in front of the bucket due to the blade thickness \( b \), is modeled using Bekker’s load–sinkage formulation~\cite{bekker1969introduction}. The pressure is given by:
\begin{equation}
P = (\frac{k_c}{b} + k_\phi) d^n,
\label{eq:P}\end{equation}
where $k_c$ and $k_\phi$ are the cohesive and frictional moduli of deformation, $n$ is the exponent of the soil deformation, and $d$ is the penetration depth, defined as $\overline{BD}$. 
With $F$ and $R$ unknown, equilibrium is enforced by force balance in the $x$ and $z$ directions; eliminating $R$ yields the expression for $F$:
\begin{equation}
\begin{split}
F & = d^2 w \gamma g N_{\gamma} + C w d N_{c}  \\
 & + C_a w d N_{a} + W_{load} N_{q},  \\
\end{split}
\label{eq:FEE}
\end{equation}
where $\gamma$ is the soil density and $W_{load}$ is the load of soil in the failure wedge. $N_{\gamma,i}$, $N_{c,i}$, $N_{a,i}$, and $N_{q,i}$ are the four bearing capacity factors given as:
\begin{align}
N_{\gamma} &= \frac{(\cot \beta - \tan \alpha)(\cos \alpha + \sin \alpha \cot (\beta + \phi))}{2[\cos (\rho + \delta) + \sin (\rho + \delta) \cot (\beta + \phi)]}, \label{eq:Ngamma}\\
N_{c} &= \frac{1 + \cot \beta \cot (\beta + \phi)}{\cos (\rho + \delta) + \sin (\rho + \delta) \cot (\beta + \phi)}, \label{eq:Nc} \\
N_{a} &= \frac{1 - \cot \rho \cot (\beta + \phi)}{\cos (\rho + \delta) + \sin (\rho + \delta) \cot (\beta + \phi)}, \label{eq:Na} \\
N_{q} &= \frac{\cos \alpha + \sin \alpha \cot (\beta + \phi)}{\cos (\rho + \delta) + \sin (\rho + \delta) \cot (\beta +\phi)},\label{eq:Nq}
\end{align}
where $\rho$ is the angle of attack. These dimensionless terms capture the effects of soil properties on excavation resistance: \(N_{\gamma}\) reflects soil self-weight, \(N_{c}\) internal cohesion, \(N_{a}\) bucket--soil adhesion, and \(N_{q}\) surcharge effects. The failure wedge angle \(\beta\) is obtained by minimizing \(N_{\gamma}\)~\cite{reece1964paper,mckyes1985_book}.

\section{Soil Parameter Identification}
\label{sec:param_identification}
The FEE-based model described in Section~\ref{sec:methodology} depends on a set of soil parameters. These parameters are represented by the vector:
\begin{equation}
\boldsymbol{\theta} = \begin{bmatrix} \gamma & C & C_a & \phi & \delta & k_c & k_{\phi} & n \end{bmatrix}^T,
\end{equation}
which are estimated from hinge-joint force measurements \(F_{H,1}\) and \(F_{H,2}\) (Fig.~\ref{fig:bucketpath}). These forces then transform into the observed tangential and normal components \(F^T_{\text{obs}}\) and \(F^N_{\text{obs}}\). All digging scenarios and force traces were generated using a validated high-fidelity wheel loader model \cite{karanfil2025developing} in Algoryx \cite{algoryx2024}, shown in Fig.~\ref{fig:Algoryx}. Algoryx models soil using DEM-based granular particles, coupled in real time with the multibody machine dynamics. The parameter estimation was carried out in sequential stages. Stage~1 targets parameters \(\theta_1 = \begin{bmatrix} C_a & \delta & k_c & k_\phi & n \end{bmatrix}^T\), affecting the tangential force \(F^T\). The total force is reconstructed from the measured normal force via \(F = F^N_{\text{obs}} / \cos \delta\) and substituted into:
\begin{equation}
F^T = w b\left(\frac{k_c}{b} + k_\phi\right) d^n + F^N_{\text{obs}} \tan \delta + C_a w L_t.
\label{eq:FT_obs}
\end{equation}
This reformulation enables an independent optimization over \( \theta_1 \) by minimizing the squared error between predicted and measured \( F^T \). The Stage~1 optimization problem is then defined as:
\begin{subequations} \label{eq:stage1}
\begin{align}
\min_{\theta_1} \;\;  J_{\theta_1} &= \sum^N_{i = 1}(F_{obs, i}^T - F_i^T)^2 \label{eq:stage1_cost} \\
\text{s.t.} \;\;  F_i^T &= w b\left(\frac{k_c}{b} + k_\phi\right) d_i^n + F_{obs,i}^N \tan \delta \nonumber \\
& + C_a w L_{t,i}, \quad i = 1, \dots, N , \label{eq:stage1_c1} \\
 \theta_{\text{1,min}} &\leq \theta_1 \leq \theta_{\text{1,max}}. \label{eq:stage1_c2}
\end{align}
\end{subequations}
The resulting solution \( \theta_1^* \) provides the parameters for Stage 2. Using the optimized external friction angle \( \delta^* \), the total force is reconstructed from the observed normal force as $F_{\text{obs}} = F^N_{\text{obs}} / \cos \delta^*$, serving as the reference to fit the parameters $\theta_2 =\begin{bmatrix} C & \gamma & \phi \end{bmatrix}^T $. The following nonlinear least-squares problem is then solved to minimize the discrepancy between observed forces and the FEE-predicted force:

\begin{subequations} \label{eq:stage2}
\begin{align}
\min_{\theta_2} \;\; J_{\theta_2} &= \sum_{i=1}^N \left( F_{\text{obs},i} - F_i \right)^2 \label{eq:stage2_cost} \\
\text{s.t.} \quad F_i &= d_i^2 w \gamma g N_{\gamma,i} + C w d_i N_{c,i} \nonumber \\
&\quad + C_a^* w d_i N_{a,i} + W_{\text{load}} N_{q,i}, \label{eq:stage2_c1} \\
N_{b,i} &= f_{N_b}(\alpha, \beta_i, \phi, \rho_i, \delta^*),\quad i = 1, \dots, N \label{eq:stage2_c2} \\
\theta_{2,\text{min}} &\leq \theta_2 \leq \theta_{2,\text{max}}. \label{eq:stage2_c3}
\end{align}
\end{subequations}
Note that \( C_a^* \) and \( \delta^* \) are fixed values obtained from Stage~1. The vector \( \mathbf{N}_{b,i} \) in~\eqref{eq:stage2_c2} represents the set of bearing capacity factors $N_{b,i} = [N_{\gamma,i}, N_{c,i}, N_{a,i}, N_{q,i}]^\top$, computed as a function of \( \alpha, \beta_i, \phi, \rho_i \), and the fixed \( \delta^* \). $f_{N_b}$ is a vector with the bearing capacity functions as in equations~\eqref{eq:Ngamma}--\eqref{eq:Nq}.

The previous two stages prioritized normal force fitting due to the reconstruction of the total force \( F \) from observed \( F^N_{\text{obs}} \). As a result, the RMSE of \( F^N \) was consistently lower than that of \( F^T \). To address this imbalance and further reduce tangential force error, we introduce a third refinement stage.

This step re-optimizes only the parameters that affect \( F^T \) but not \( F^N \), identified by analyzing equations~\eqref{eq:FT} and~\eqref{eq:FN}. The compaction pressure term \( P \), modeled via Bekker’s formulation \cite{bekker1969introduction}, influences only \( F^T \), leading to the reduced parameter set $\theta_3 = \begin{bmatrix}  k_c & k_\phi & n \end{bmatrix}^T$:
\begin{subequations} \label{eq:stage3}
\begin{align}
\min_{\theta_3} \;\;   J_{\theta_3} &= \sum_{i=1}^{N} (F^T_{\text{obs},i} - F^T_i)^2 \label{eq:stage3_cost}\\
\text{s.t.} \;\; F_i^T & = w b P_i + F_i \sin \delta^* + C_a^* w L_{t,i}, \label{eq:stage3_c1} \\
F^N_i & = F_i \cos \delta^*, \label{eq:stage3_c2} \\
 P_i &= \left(\frac{k_c}{b} + k_\phi\right) d_i^n, \label{eq:stage3_c3} \\
F_i  &= d_i^2 w \gamma^* g N_{\gamma,i} + C w d_i N_{c,i} \nonumber \\
& + C_a^* w d_i N_{a,i} +W_{\text{load}} N_{q,i}, 
 \label{eq:stage3_c4} \\
N_{b,i} &= f_{N_b}(\alpha, \beta_i, \phi^*, \rho_i, \delta^*),\quad i = 1, \dots, N \label{eq:GD_baseline_c5} \\
\theta_{3,\text{min}} & \leq \theta_3 \leq \theta_{3,\text{max}} .\label{eq:GD_baseline_c9}
\end{align}
\end{subequations}
To ensure physically meaningful results, we imposed hard constraints on the soil parameters based on representative values from prior work~\cite{abdolmohammadi2025data} and the Algoryx library~\cite{algoryx2024}. The density was constrained as 
\(\gamma \in [1200,\,2500]~\mathrm{kg/m^3}\); cohesion and adhesion as 
\(C, C_a \in [0,\,2500]~\mathrm{N/m^2}\); and friction angles as 
\(\phi, \delta \in [0,\,0.785]~\mathrm{rad}\). For Bekker parameters, we set 
\(k_c \in [0,100]~\mathrm{N/m^{n+1}}, k_\phi \in [0,5000]~\mathrm{kN/m^{n+2}}, 
n \in [0.1,\,1.5]\). These ranges are consistent with typical soil mechanics data reported for granular and cohesive soils \cite{look2007handbook}.

\subsection{Parameter Sensitivity Analysis and Reduced-Order Model}
\label{sec:param_sens}

The proposed parameter-optimization algorithm improves calibration efficiency, achieving up to a 50\% reduction in convergence time compared to single-step full model optimization \cite{abdolmohammadi2025data}. Yet, the high dimensionality of the parameter space raises the question of which parameters most strongly influence the model. A global sensitivity analysis was performed to quantify each parameter’s impact on excavation force and identify low-influence ones for reduced-order models to decrease the dimensionality of the problem. The resultant force $F_R$ was computed as the Euclidean norm of tangential and normal force components $F_R = \sqrt{(F^T)^2+(F^N)^2}$, and Sobol’s method \cite{sobol2001global} was adopted. This global approach, unlike local derivative-based techniques, captures nonlinearities and parameter interactions, with the total-order Sobol index ($S_T$) used to measure each parameter’s overall contribution to output variance.

The input parameter ranges required for this analysis are the same explained in Section \ref{sec:param_identification}. To visualize sensitivity results over several orders of magnitude, we use a logarithmic scale on the vertical axis of the bar chart. This is necessary due to several parameters, such as $\gamma$, $\phi$, $\delta$, $C$, and $C_a$, exhibiting very low sensitivity indices (on the order of $10^{-4}$ to $10^{-12}$) compared to the other parameters.

\begin{figure}
    \centering
    \includegraphics[width=0.96\linewidth]{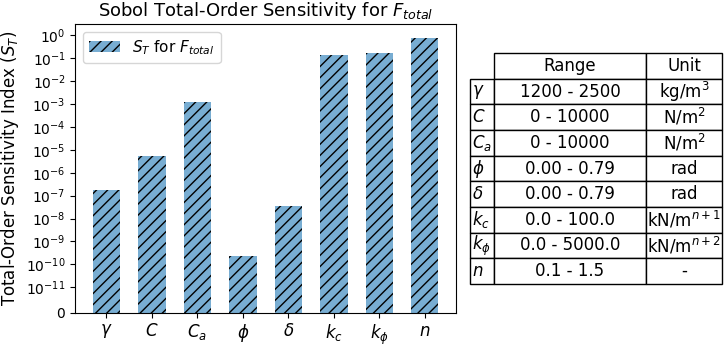}
    \caption{ Sobol total-order sensitivities (\(S_T\)) of resultant force \(F_R\). The bar chart shows parameter influence on a logarithmic scale; parameter ranges and units are listed in the table.}
    \label{fig:sobol}
\end{figure}

The results in Fig.~\ref{fig:sobol} show the ranked sensitivity of parameters in \( F_R \). Among all parameters, \( n \) exhibits the highest influence, followed by \( k_c \) and \( k_\phi \). Next, \( C_a \) shows a significantly lower influence, followed by \( C \), \( \gamma \), \( \delta \), and \( \phi \) in decreasing order. This ordering highlights the opportunity for dimensionality reduction by fixing low-impact parameters, such as \( \phi \), \( \delta \), and \( \gamma \).

\subsection{Reduced Order Soil Parameter Identification Results}
\begin{figure*}[t]
    \begin{subfigure}{0.23\textwidth}
        \centering
        \includegraphics[width=\linewidth]{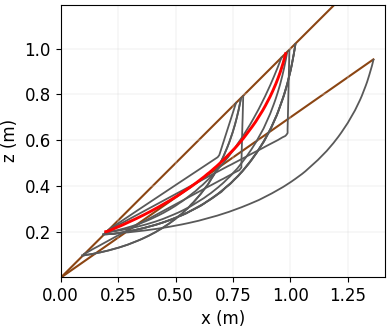}
        \caption{}
        \label{fig:bucket_paths}
    \end{subfigure}
        \begin{subfigure}{0.44\textwidth}
        \centering
        \includegraphics[width=\textwidth]{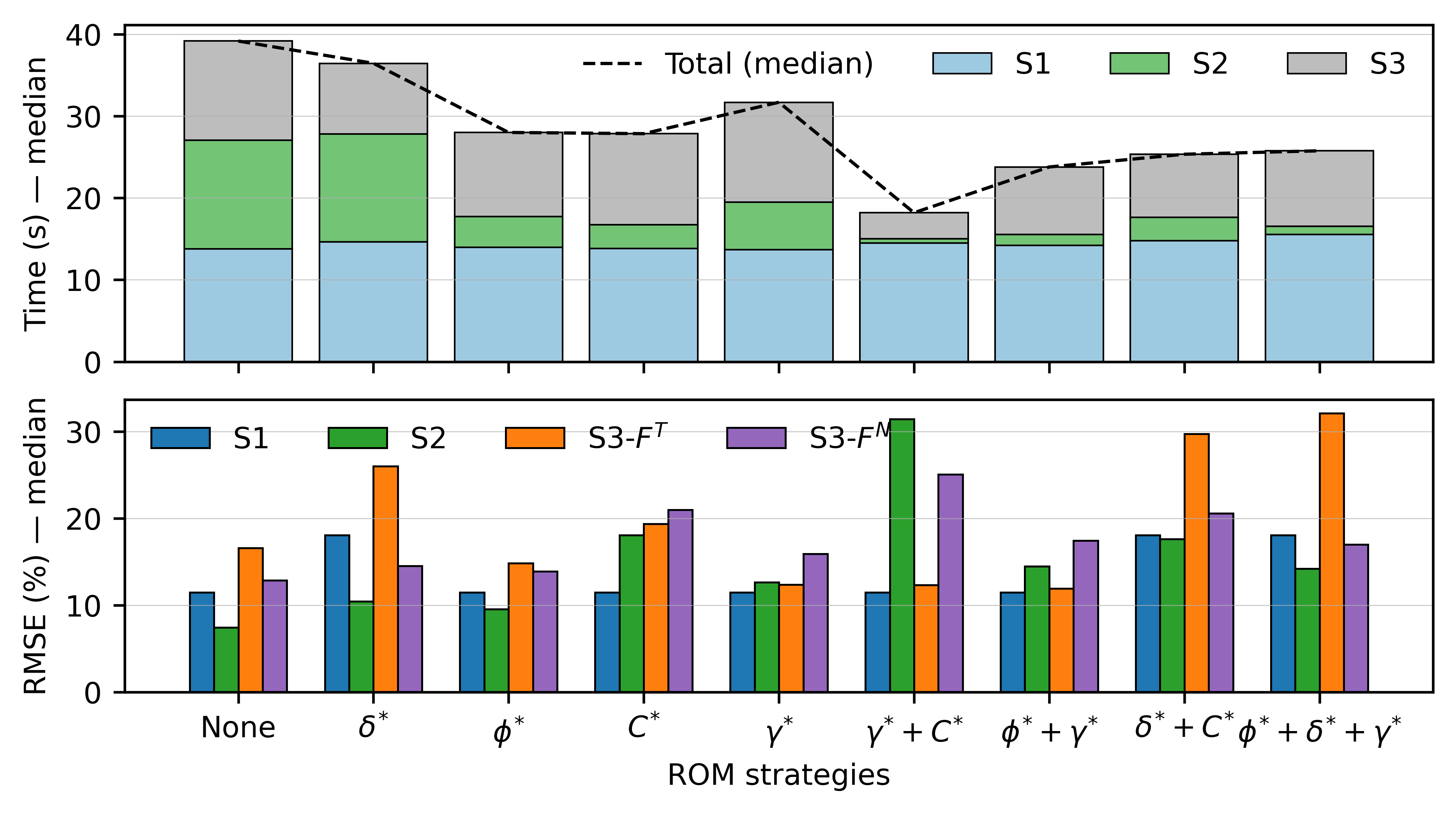}
        \caption{}
        \label{fig:romstrategies}
    \end{subfigure}
    \begin{subfigure}{0.33\textwidth}
        \centering
        \includegraphics[width=\linewidth]{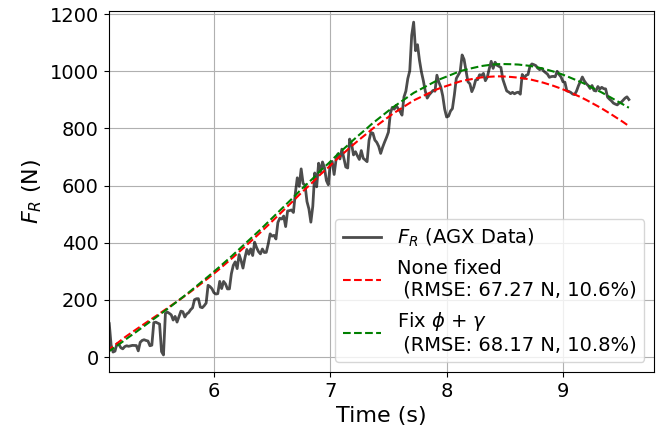}
        \caption{}
        \label{fig:forces}
    \end{subfigure}
    \caption{(a) Bucket-tip trajectories in the $x$–$z$ plane with soil edges at $35^\circ$ and $45^\circ$. (b) ROM performance: median convergence times (top) and median RMSE for $F^T$ and $F^N$ (bottom). * indicates fixed parameters and S1,S2 and S3 refers to the stage 1, 2, and 3 of the parameter identification problem. (c) $F_R$ for one path for full order ROM and ROM with fixed $\phi^*+\gamma^*$.}
\end{figure*}
Guided by Sobol ranking, we formed 28 reduced–order models (ROMs) by fixing subsets of the five low-influence parameters \((\phi,\delta,\gamma,C,C_a)\). Each ROM was solved with the three–stage identification under two regimes: repeated runs (identical initialization) and randomized starts (12 initial guesses). Repeated runs yielded effectively deterministic outcomes per ROM (boxplots collapse to the median), indicating negligible intrinsic solver variance. With randomized starts, performance remained robust for most ROMs; wider spreads occurred primarily when the external friction \(\delta\) was fixed, either alone or in combinations, suggesting multi-modal behavior tied to \(\delta\). It was also observed that the variance of the total execution time decreased as additional parameters were held fixed.

To test generality, we evaluated nine bucket paths over \(35^\circ/45^\circ\) piles (smooth and piecewise-linear) across three soils (gravel, sand, dirt) in simulation as shown in Fig~\ref{fig:bucket_paths}. Fig~\ref{fig:romstrategies} reports medians across selected ROM strategies. Fixing \(\phi^*\) or \(C^*\) reduced total median time from \(\sim39\,\mathrm{s}\) (full model) to \(\sim28\text{–}30\,\mathrm{s}\) with little change in stage balance and minor RMSE impact. \(\gamma^*\) offered smaller time savings. The pair \(\gamma^*{+}C^*\) produced the shortest runs (\(\sim18\,\mathrm{s}\)) but increased Stage-3 (S3) error; strategies containing \(\delta^*\) similarly inflated S3 RMSE without time gains. Overall, \(\phi^*\) is a safe fix; \(C^*\) or \(\gamma^*\) are acceptable when modest RMSE growth is tolerable. Nominated best result $\phi^*+\gamma^*$ showed a 27\% time saving (from 40 s down to 29 s). For this ROM strategy, for a single digging path shown as red in Fig~\ref{fig:bucket_paths}, the simulation forces and estimated forces are shown in Fig~\ref{fig:forces}.

\section{Optimal Bucket Path Planner}

\begin{figure*}
    \centering
    \includegraphics[width=1\linewidth]{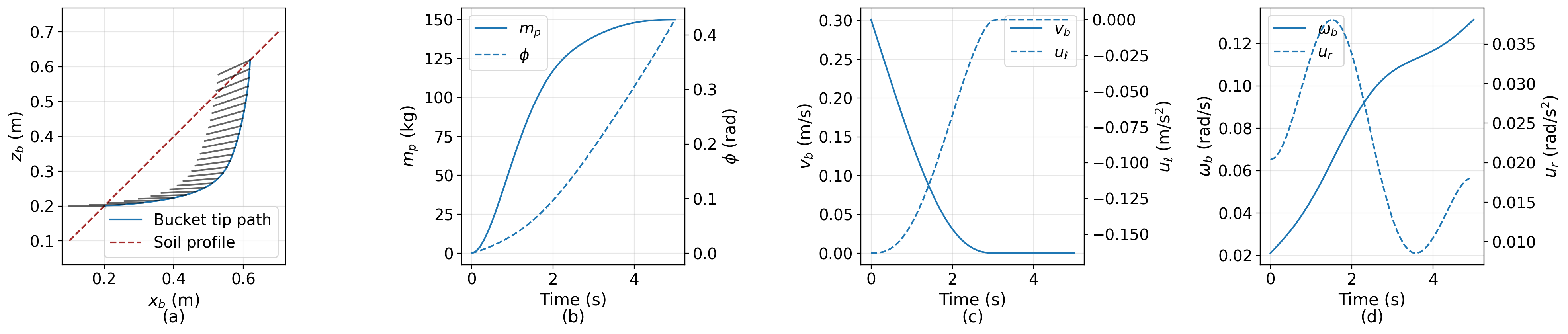}
    \caption{Results of the model predictive path planner.
    (a) Optimized bucket tip trajectory in the $x$--$z$ plane with soil profile and bucket orientation indicators. 
    (b) Accumulated payload mass $m_p$ and bucket angle $\phi$ over time. 
    (c) Bucket tip velocity $v_b$ and corresponding input $u_\ell$. 
    (d) Angular velocity $\omega_b$ and corresponding input $u_r$.}
    \label{fig:mppp_results}
\end{figure*}
With the soil parameters identified, bucket motion is planned using the FEE model. Prior work minimized fuel consumption through an engine–hydraulic formulation with FEE interactions~\cite{yao2023bucket}, whereas this work directly minimizes excavation energy at the bucket level, yielding a lower-dimensional problem with faster convergence.

Bucket motion is described in a $T$–$N$ frame fixed at the tip as shown in Fig~\ref{fig:MPPP}. The tip position is $(x_b, z_b)$ in the global $x$–$z$ plane with orientation $\Phi$. The soil center of mass lies at a fixed distance $r$ from the tip, yielding tip velocity components $v_b$ (tangential) and $r\omega_b$ (rotational), where $\omega_b$ is the angular velocity. The angle between them, $\Theta$, is assumed constant under a quasi-static approximation. The bucket's dynamics can be described by the following set of first-order differential equations:
\begin{equation}
\dot{\Phi}(t) = \omega_b(t), \quad \dot{\omega}_b(t) = u_r(t), \quad \dot{v}_b(t) = u_l(t).
\end{equation}
Here, \( u_l \) (m/s\(^2\)) represents the linear acceleration of the bucket tip, and \( u_r \) (rad/s\(^2\)) denotes the angular acceleration of the bucket about its rotation axis. These two control inputs are assumed to be executable by a lower-level controller, and the effort required to realize them is delegated to the design of the path tracking controller. To compute the motion of the bucket tip in the global coordinate frame, a transformation from the local tangential–normal $T–N$ frame to the global $x-z$ frame is applied. The resulting velocity components in global coordinates are given by:
\begin{align}
\dot{x}_b &= ( v_b + r\omega_b \cos \Theta) \cos \Phi - r\omega_b \sin \Theta \sin \Phi, \label{eq:xdot} \\
\dot{z}_b &= ( v_b + r\omega_b \cos \Theta) \sin \Phi + r\omega_b \sin \Theta \cos \Phi. \label{eq:zdot} 
\end{align}
 To incorporate soil collection into trajectory optimization, the bucket mass is tracked as a dynamic state to enforce payload constraints within the horizon. Soil accumulation is modeled as:
 \begin{equation}
     \dot{m}_p(t) = \gamma \, \omega \, \dot{A}(t),
 \end{equation}
 where $\dot{A} = \dot{x}_b (p(x_b)-z_b)$ is the rate of change of cross-sectional area under the pile profile $p(x)$. 

By incorporating the forces acting at the bucket tip, computed using the FEE model described in Section~\ref{sec:methodology}, the instantaneous power required during the bucket-loading cycle can be estimated at each time step. The total required power \( P_r \) is expressed as:
\begin{align}
P_r &= ( F^T + F_B \sin \Phi)( v_b + r\omega_b \cos \Theta) \nonumber \\
    &+ \left( F^N + F_B \cos \Phi \right)\left( r\omega_b \sin \Theta \right) \label{eq:Pr},
\end{align}
where $F_B$ is the force due to bucket weight and $F^T$ and $F^N$ come from the FEE explained in Section \ref{sec:methodology}. The bucket-tip dynamics are then expressed in state–space form, with states including soil mass, tip kinematics, and pose, and inputs given by linear and angular accelerations.
\begin{equation}
\mathbf{x} = \begin{bmatrix}
 m_p &v_b & \omega_b & \Phi & x_b & z_b
\end{bmatrix}^\top
\quad
\mathbf{u} = \begin{bmatrix}
    u_l & u_r
\end{bmatrix}^\top.
\end{equation}
The dynamics are written in control-affine form,
\begin{equation}
\dot{\mathbf{x}} = f(\mathbf{x}) + B \mathbf{u},
\end{equation}
where \(f(x)\) denotes the drift term capturing mass-accumulation and kinematic relationships computed from (14) to (17), and \(B\) is the input matrix that maps the control inputs \(u\) to the state dynamics.

We formulate the nonlinear optimal problem path planning over a horizon of \(n\) steps with uniform step size \(\Delta T\), and the final time \(t_f=n\,\Delta T\), as follows:

\begin{subequations} \label{eq:MPPP}
\begin{align}
\min \;\;   J &= \sum_{i=1}^{N} P_r(x_i)^2 +\sum_{k=1}^{N-1} \lambda_{\dot u} \left\lVert \frac{u_k - u_{k-1}}{\Delta T} \right\rVert_2^2 , \label{eq:MPPP_cost} \\
\text{s.t.} \;\; x_{i+1} & = x_i + \frac{\Delta T}{6} \left( k_1 + 2k_2 + 2k_3 + k_4 \right), \nonumber \\
 &\quad\quad\quad\quad\quad i = 1, \dots, N, \label{eq:MPPP_c1} \\
x_{min} &\leq x \leq x_{max}, \label{eq:MPPP_c2} \\
u_{min} &\leq u \leq u_{max}, \label{eq:MPPP_c3} \\
x(0) &= x_0,  \label{eq:MPPP_c5} \\
z_b(t_f) & = p(x_b(t_f)), \label{eq:MPPP_c6} \\
m_{p,f} &\geq m_{min}. \label{eq:MPPP_c7} 
\end{align}
\end{subequations}
The objective function in \eqref{eq:MPPP_cost} minimizes the normalized instantaneous power \(P_{r,i}\) delivered during excavation, where it is evaluated from the FEE model at each discretization point \(i\) based on \eqref{eq:Pr}. An additional term is added to the cost function to penalize the rate of change of the inputs tuned by $\lambda_{\dot u}$ to promote smooth inputs. The dynamics constraint \eqref{eq:MPPP_c1} enforces the continuous–time model \(\dot{\mathbf{x}}=f(\mathbf{x})+B\,\mathbf{u}\) using a fourth–order Runge–Kutta (RK4) scheme, where $k_1, k_2, k_3$ and $k_4$ denote the Runge–Kutta coefficients. The constraints \eqref{eq:MPPP_c2}–\eqref{eq:MPPP_c3} bound states and inputs within admissible sets, workspace limits, and velocity/acceleration caps. The initial condition \eqref{eq:MPPP_c5} fixes the optimizer to the state at the beginning of the horizon. The terminal equality \eqref{eq:MPPP_c6} imposes a desired exit condition by constraining the tip to lie on the stockpile curve at \(t_f\). Finally, the payload requirement \eqref{eq:MPPP_c7} enforces production by requiring the final collected mass \(m_{p,f}\) at \(t_f\) to meet or exceed a specified minimum payload.

The decision variables are the full state trajectory 
\(\{x_i\}_{i=0}^{N}\) and input sequence \(\{u_i\}_{i=0}^{N-1}\). The problem is formulated as a single Nonlinear Program (NLP) that optimizes the entire bucket trajectory over the full horizon within a single optimization problem.

\section{Results}
To evaluate the proposed optimal path planner, we implemented the framework using the soil parameters identified through a previous digging cycle highlighted in red in Fig~\ref{fig:bucket_paths} and the parameter optimization procedure with the selected ROM strategy ($\phi +\gamma$ fixed). Table~\ref{tab:params} summarizes calibrated soil parameters as well as the geometric and operating conditions of the environment and wheel loader, namely the bucket width $w$, cutting edge width $b$, soil angle $\alpha$, and angle $\Theta$ between $v_b$ and $r\omega_b$. The model predictive path planner was formulated in CasADi \cite{CASADI} with the FEE model embedded into the stage cost to penalize instantaneous power demand, and the nonlinear program was solved with IPOPT \cite{IPOPT}. All runs were initialized from the same guess and executed on the same workstation (14th-gen Intel Core i9). 

\begin{table}[h!]
\centering
\caption{Parameters for the \textit{FEE} Model and Planner}
\begin{tabular}{p{1.8cm} p{0.5cm} p{1.8cm} p{0.5cm} p{1.2cm} p{0.5cm}}
\hline
\multicolumn{6}{c}{\textbf{Soil Parameters}} \\
\hline
Parameter & Value & Parameter & Value & Parameter & Value \\
\hline
$k_c$ (kN/m$^{n+1}$)     & 1.93   & $k_\phi$ (kN/m$^{n+2}$) & 0.19   & $n$ (--)    & 0.94 \\
$C_a$ (N/m$^2$)& 53.8  & $C$ (N/m$^2$)      & 518   & $\gamma$ (kg/m$^3$) & 1850 \\
$\delta$ (rad)& 0.609 & $\phi$ (rad) & 0.075 &  &  \\
\hline
\multicolumn{6}{c}{\textbf{Environment and Wheel Loader Parameters}} \\
\hline
$w$ (m)       & 1.69  & $b$ (m)      & 0.03  & $\alpha$ (rad) & 0.785 \\
$\Theta$ (rad) & 1.099 &  &  &  &  \\
\hline
\end{tabular}
\label{tab:params}
\end{table}

To formulate the optimal path planner, we enforced bounds on both the states and the control inputs.  
The bucket tip position was restricted within $0 \leq x_b, z_b \leq 1.0~\text{m}$, while the payload mass $m_p$ was constrained to remain nonnegative.  
The bucket tip velocity was constrained to $0 \leq v_b \leq 1.0~\text{m/s}$, the angular velocity to $-1.0 \leq \omega_b \leq 1.0~\text{rad/s}$, and the bucket angle to $0 \leq \phi \leq 0.69$ rad.
The control inputs were bounded in a similar range, with $-1.0 \leq u_\ell \leq 1.0~\text{m/s}^2$ for linear acceleration and $-1.0 \leq u_r \leq 1.0~\text{rad/s}^2$ for angular acceleration, inspired by~\cite{yao2023bucket}. The discretized horizon was set to $N=50$ with a sampling interval of $\Delta T = 100ms$, corresponding to a total problem horizon of $5s$. A minimum payload threshold of $m_{min} = 150 kg$ was imposed to ensure that the bucket achieved sufficient filling by the end of the trajectory. The initial state vector was specified as  $x_0 = [0, 0.1, 0, 0, 0.1, 0.1]^\top$, representing an empty bucket positioned at the soil face, with a small initial forward velocity, zero angular velocity, and zero angle of attack. Finally, the tuning parameters were set to $\lambda_{\dot u} = 1\mathrm{e}{-4}$ for the input-rate penalty.
 
 The nonlinear program converged in a total solve time of approximately 1.2 seconds. The optimization results are illustrated in Fig.~\ref{fig:mppp_results}. Fig.~\ref{fig:mppp_results}(a) shows the optimal bucket tip trajectory in the $x_b$--$z_b$ plane, together with the soil profile. The path smoothly penetrates the pile, maintaining contact below the free surface, while the red orientation markers depict the evolution of the bucket angle. Fig.~\ref{fig:mppp_results}(b) shows the payload mass $m_p$, which increases monotonically as the bucket advances into the pile, and the bucket angle $\phi$, which steadily rises toward the final value of approximately $0.35~\text{rad}$. Fig.~\ref{fig:mppp_results}(c) presents the bucket tip velocity $v_b$ together with its input $u_l$, showing an initial acceleration phase followed by deceleration as the bucket becomes loaded. Fig.~\ref{fig:mppp_results}(d) shows the angular velocity $\omega_b$ and its input $u_r$, where the angular actuation regulates bucket rotation in coordination with penetration to achieve efficient filling.

\begin{figure}
    \centering
    \includegraphics[width=1\linewidth]{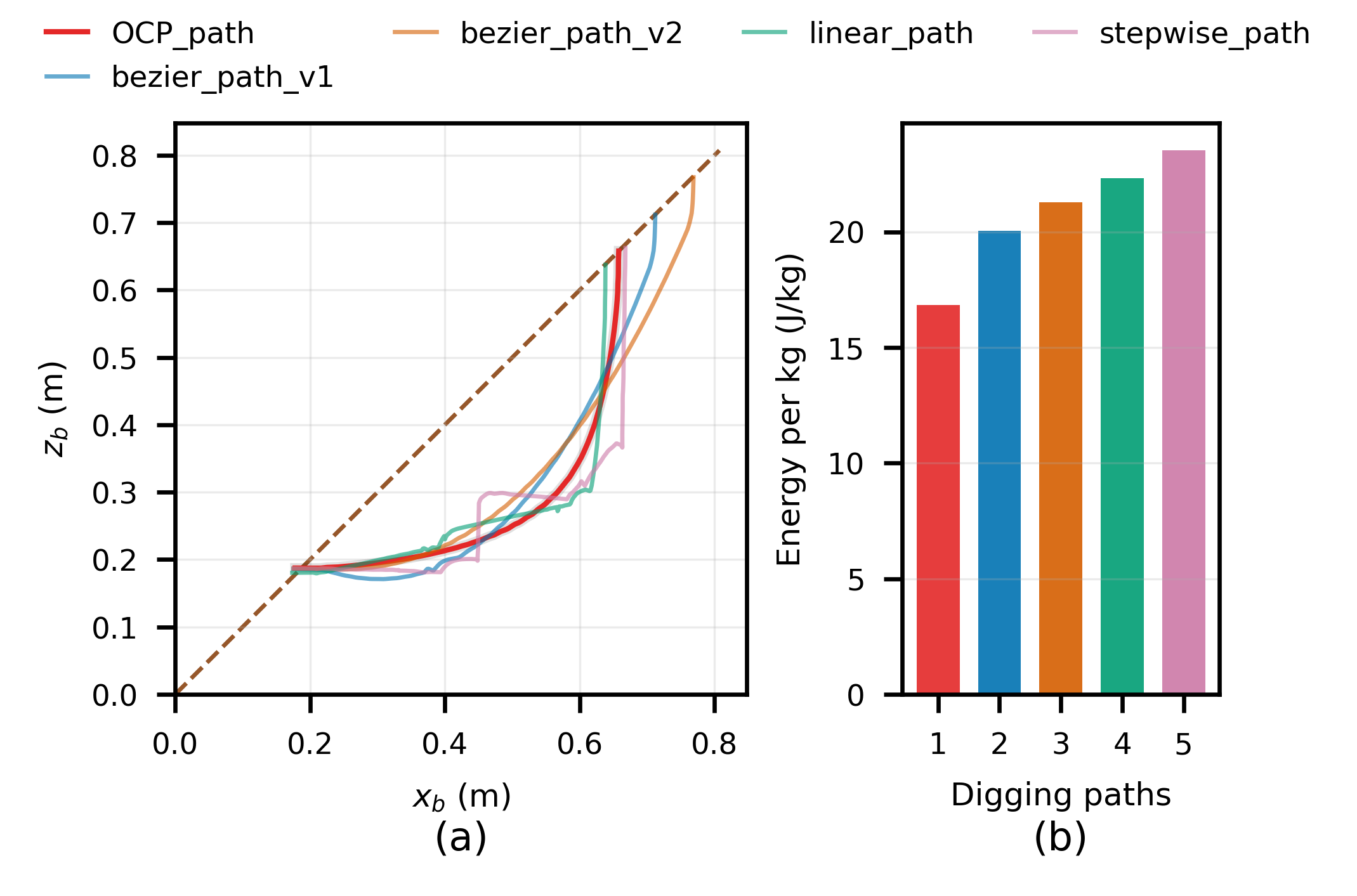}
    \caption{Validation in Algoryx: (a) bucket-tip trajectories in the $x-z$ plane with the soil line. (b) energy per picked up pass $(J/kg)$ for similar digging paths.}
    \label{fig:agx_validation}
\end{figure}

The generated optimal bucket path was validated in Algoryx by simulating its trajectory alongside matched comparison paths with identical start conditions and enclosed area to ensure equal payload. The set included smooth Bézier paths, a two-segment linear path, and a stepwise path (Fig.~\ref{fig:agx_validation}a). The stepwise path is inspired by the observation from operators where they drive straight until resistance rises, then lift and repeat.

Due to uncertainties and effects such as spillage, the actual collected masses differed 
despite being designed for equal pickup. With a $150\,\text{kg}$ target, the optimal bucket path collected 
$\sim130\,\text{kg}$, the smooth Bézier paths $\sim125$–$135\,\text{kg}$, the linear path $\sim115\,\text{kg}$ (lowest), and the stepwise path $\sim155\,\text{kg}$ (highest), showing that operator-inspired strategies can increase fill factor. For a fair comparison, efficiency was measured as energy per kilogram of soil (Fig.~\ref{fig:agx_validation}b). Optimal bucket path required the least at $\sim17\,\text{J/kg}$, compared to $20$–$21\,\text{J/kg}$ (+15–25\%) for Bézier, $22$–$23\,\text{J/kg}$ (+30\%) for linear, and $24$–$25\,\text{J/kg}$ (+40\%) for stepwise. Although scenario-dependent, these results show the optimal bucket path delivers the lowest energy 
per cycle across these representative bucket-loading styles.

\section{Conclusion}
This work introduces an efficient framework for automating wheel loader operations by combining an analytical soil–tool interaction model with nonlinear parameter estimation and optimal control. The approach achieves accurate excavation force prediction and energy-optimal bucket trajectory planning, with a total computation time of about 30 seconds, comparable to a single short loading cycle. Validation using a high-fidelity model confirms prediction accuracy and demonstrates 15-40\% energy savings over standard paths. These findings highlight the feasibility of real-time, model-based optimization for earthmoving automation. Future efforts will focus on extending the framework to optimize complete wheel loader cycles, further improving productivity and energy efficiency in operations.
\bibliography{References}

\begin{thebibliography}{10}
\providecommand{\url}[1]{#1}
\csname url@samestyle\endcsname
\providecommand{\newblock}{\relax}
\providecommand{\bibinfo}[2]{#2}
\providecommand{\BIBentrySTDinterwordspacing}{\spaceskip=0pt\relax}
\providecommand{\BIBentryALTinterwordstretchfactor}{4}
\providecommand{\BIBentryALTinterwordspacing}{\spaceskip=\fontdimen2\font plus
\BIBentryALTinterwordstretchfactor\fontdimen3\font minus \fontdimen4\font\relax}
\providecommand{\BIBforeignlanguage}[2]{{%
\expandafter\ifx\csname l@#1\endcsname\relax
\typeout{** WARNING: IEEEtran.bst: No hyphenation pattern has been}%
\typeout{** loaded for the language `#1'. Using the pattern for}%
\typeout{** the default language instead.}%
\else
\language=\csname l@#1\endcsname
\fi
#2}}
\providecommand{\BIBdecl}{\relax}
\BIBdecl

\bibitem{gottschalk2018test}
M.~Gottschalk, G.~Jacobs, and A.~Kramer, ``Test method for evaluating the energy efficiency of wheel loaders,'' \emph{ATZoffhighway worldwide}, vol.~11, pp. 44--49, 2018.

\bibitem{haas2025combined}
M.~Haas, A.~Abdolmohammadi, and S.~Nazari, ``Combined control and design optimization of a parallel electric-hydraulic hybrid wheel loader to prolong battery lifetime,'' \emph{Authorea Preprints}, 2025.

\bibitem{filla2014_compare_trajectory}
R.~Filla, M.~Obermayr, and B.~Frank, ``A study to compare trajectory generation algorithms for automatic bucket filling in wheel loaders,'' in \emph{3rd Commercial Vehicle Technology Symposium}, 2014, pp. 588--605.

\bibitem{FRANK2018}
\BIBentryALTinterwordspacing
B.~Frank, J.~Kleinert, and R.~Filla, ``Optimal control of wheel loader actuators in gravel applications,'' \emph{Automation in Construction}, vol.~91, pp. 1--14, 2018. [Online]. Available: \url{https://www.sciencedirect.com/science/article/pii/S0926580517308002}
\BIBentrySTDinterwordspacing

\bibitem{Aoshima_2021}
\BIBentryALTinterwordspacing
K.~Aoshima, M.~Servin, and E.~Wadbro, ``Simulation-based optimization of high-performance wheel loading,'' 2021. [Online]. Available: \url{https://arxiv.org/abs/2107.14615}
\BIBentrySTDinterwordspacing

\bibitem{chen2022}
Y.~Chen, H.~Jiang, G.~Shi, and T.~Zheng, ``Research on the trajectory and operational performance of wheel loader automatic shoveling,'' \emph{Applied sciences}, vol.~12, no.~24, p. 12919, 2022.

\bibitem{halbach2019_Pile_Loading_Controller}
E.~Halbach, J.~K{\"a}m{\"a}r{\"a}inen, and R.~Ghabcheloo, ``Neural network pile loading controller trained by demonstration,'' in \emph{2019 International Conference on Robotics and Automation (ICRA)}.\hskip 1em plus 0.5em minus 0.4em\relax IEEE, 2019, pp. 980--986.

\bibitem{Eriksson2024}
D.~Eriksson, R.~Ghabcheloo, and M.~Geimer, ``Automatic loading of unknown material with a wheel loader using reinforcement learning,'' in \emph{2024 IEEE International Conference on Robotics and Automation (ICRA)}, 2024, pp. 3646--3652.

\bibitem{azulay2021_scooping_control}
O.~Azulay and A.~Shapiro, ``Wheel loader scooping controller using deep reinforcement learning,'' \emph{IEEE access}, vol.~9, pp. 24\,145--24\,154, 2021.

\bibitem{reece1964paper}
A.~Reece, ``Paper 2: The fundamental equation of earth-moving mechanics,'' in \emph{Proceedings of the institution of mechanical engineers, conference proceedings}, vol. 179, no.~6.\hskip 1em plus 0.5em minus 0.4em\relax SAGE Publications Sage UK: London, England, 1964, pp. 16--22.

\bibitem{mckyes1985_book}
E.~McKyes, \emph{Soil cutting and tillage}.\hskip 1em plus 0.5em minus 0.4em\relax Elsevier, 1985.

\bibitem{yao2023bucket}
J.~Yao, C.~P. Edson, S.~Yu, G.~Zhao, Z.~Sun, X.~Song, and K.~A. Stelson, ``Bucket loading trajectory optimization for the automated wheel loader,'' \emph{IEEE transactions on vehicular technology}, vol.~72, no.~6, pp. 6948--6958, 2023.

\bibitem{egli2022soil}
P.~Egli, D.~Gaschen, S.~Kerscher, D.~Jud, and M.~Hutter, ``Soil-adaptive excavation using reinforcement learning,'' \emph{IEEE robotics and automation letters}, vol.~7, no.~4, pp. 9778--9785, 2022.

\bibitem{yu2023line}
S.~Yu, X.~Song, and Z.~Sun, ``On-line prediction of resistant force during soil--tool interaction,'' \emph{Journal of Dynamic Systems, Measurement, and Control}, vol. 145, no.~8, p. 081004, 2023.

\bibitem{tan2005online}
C.~P. Tan, Y.~H. Zweiri, K.~Althoefer, and L.~D. Seneviratne, ``Online soil parameter estimation scheme based on newton-raphson method for autonomous excavation,'' \emph{IEEE/ASME Transactions on Mechatronics}, vol.~10, no.~2, pp. 221--229, 2005.

\bibitem{abdolmohammadi2025data}
A.~Abdolmohammadi, N.~Mojahed, S.~Nazari, and B.~Ravani, ``Data-efficient excavation force estimation for wheel loaders,'' \emph{arXiv preprint arXiv:2506.22579}, 2025.

\bibitem{algoryx2024}
\BIBentryALTinterwordspacing
A.~Simulations, ``Agx dynamics,'' September 2024, accessed: 2024-09-26. [Online]. Available: \url{https://www.algoryx.se/products/agx-dynamics/}
\BIBentrySTDinterwordspacing

\bibitem{karanfil2025developing}
D.~Karanfil, D.~Lindmark, M.~Servin, D.~Torick, and B.~Ravani, ``Developing a calibrated physics-based digital twin for construction vehicles,'' \emph{arXiv preprint arXiv:2508.08576}, 2025.

\bibitem{abtahi2025multi}
M.~Abtahi, M.~Rabbani, A.~Abdolmohammadi, and S.~Nazari, ``Multi-step deep koopman network (mdk-net) for vehicle control in frenet frame,'' \emph{arXiv preprint arXiv:2503.03002}, 2025.

\bibitem{corke2000}
P.~Corke, J.~Trevelyan, H.~Cannon, and S.~Singh, ``Models for automated earthmoving,'' in \emph{Experimental Robotics VI}.\hskip 1em plus 0.5em minus 0.4em\relax Springer, 2000, pp. 163--172.

\bibitem{worley2008}
M.~Worley and V.~La~Saponara, ``A simplified dynamic model for front-end loader design,'' \emph{Proceedings of the Institution of Mechanical Engineers, Part C: Journal of Mechanical Engineering Science}, vol. 222, no.~11, pp. 2231--2249, 2008.

\bibitem{bekker1969introduction}
M.~G. Bekker, \emph{Introduction to terrain-vehicle systems}.\hskip 1em plus 0.5em minus 0.4em\relax The University of Michigan Press, 1969.

\bibitem{look2007handbook}
B.~G. Look, \emph{Handbook of geotechnical investigation and design tables}.\hskip 1em plus 0.5em minus 0.4em\relax Taylor \& Francis, 2007.

\bibitem{sobol2001global}
I.~M. Sobol, ``Global sensitivity indices for nonlinear mathematical models and their monte carlo estimates,'' \emph{Mathematics and computers in simulation}, vol.~55, no. 1-3, pp. 271--280, 2001.

\bibitem{CASADI}
J.~A. Andersson, J.~Gillis, G.~Horn, J.~B. Rawlings, and M.~Diehl, ``Casadi: a software framework for nonlinear optimization and optimal control,'' \emph{Mathematical Programming Computation}, vol.~11, pp. 1--36, 2019.

\bibitem{IPOPT}
A.~W{\"a}chter and L.~T. Biegler, ``On the implementation of an interior-point filter line-search algorithm for large-scale nonlinear programming,'' \emph{Mathematical programming}, vol. 106, pp. 25--57, 2006.

\end{thebibliography}

\end{document}